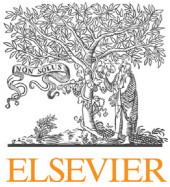

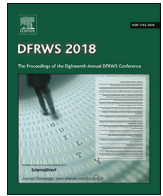

DFRWS 2018 USA — Proceedings of the Eighteenth Annual DFRWS USA

# Deep learning at the shallow end: Malware classification for non-domain experts

Quan Le [a, *], Oisín Boydell [a], Brian Mac Namee [a], Mark Scanlon [b]

[a] Centre for Applied Data Analytics Research, University College Dublin, Ireland
[b] Forensics and Security Research Group, University College Dublin, Ireland

Keywords:
Deep learning
Machine learning
Malware analysis
Reverse engineering

## ABSTRACT

Current malware detection and classification approaches generally rely on time consuming and knowledge intensive processes to extract patterns (signatures) and behaviors from malware, which are then used for identification. Moreover, these signatures are often limited to local, contiguous sequences within the data whilst ignoring their context in relation to each other and throughout the malware file as a whole. We present a Deep Learning based malware classification approach that requires no expert domain knowledge and is based on a purely data driven approach for complex pattern and feature identification.

© 2018 The Author(s). Published by Elsevier Ltd on behalf of DFRWS. This is an open access article under the CC BY-NC-ND license (http://creativecommons.org/licenses/by-nc-nd/4.0/).

## 1. Introduction

In law enforcement agencies throughout the world, there are growing digital forensic backlogs of unimaged, unprocessed, and unanalyzed digital devices stored in evidence lockers (Scanlon, 2016). This growth is attributable to several compounding factors. The sheer volume of cases requiring digital forensic processing extends far beyond digitally executed crimes such as phishing, online sharing of illicit content, online credit card fraud, etc., to "traditional" crimes such as murder, stalking, financial fraud, etc. The volume of data to be analyzed per case is continuously growing and there is a limited supply of trained personnel capable of the expert, court-admissible, reproducible analysis that digital forensic processing requires.

In order to address the latter factor, many police forces have been implementing a first responder/triage model to enable on-site evidence seizure and securing the integrity of the evidence gathered (Hitchcock et al., 2016). These models train field officers in the proficient handling of digital devices at a crime scene enabling the available expert digital investigators to remain in the laboratory processing cases. In this model, the first responders are not trained in the analysis or investigation phase of the case, but can ensure the integrity and court-admissibility of the gathered evidence.

While physical resourcing in terms of hardware, training first responders, and increased numbers of expertly skilled personnel can increase an agency's digital forensic capacity, the digital forensic research community has identified the need for automation and intelligent evidence processing (Sun, 2010). One of the more labor intensive and highly-skilled tasks encountered in digital forensic investigation is malware analysis. A common technique for analyzing malware is to execute the malware in a sandbox/virtual machine to gain insight to the attack vector, payload installation, network communications, and behavioral analysis of the software with multiple snapshots taken throughout the analysis of the malware lifecycle. This is an arduous, time-consuming, manual task that can often span over several days. A survey of digital forensic examiners conducted by Hibshi et al. (2011) found that users are often overwhelmed by the amount of technical background required to use common forensic tools. This results in a high barrier to entry for digital investigators to expand their skillset to incorporate additional topics of expertise, such as malware analysis.

Artificial Intelligence (AI) combined with automation of digital evidence processing at appropriate stages of an investigation has significant potential to aid digital investigators. AI can expedite the investigative process and ultimately reduce case backlog while avoiding bias and prejudice (James and Gladyshev, 2013). Overviews of the applications of AI to security and digital forensics are provided in (Franke and Srihari, 2008) and (Mitchell, 2014). A number of approaches have been implemented to aid digital forensic investigation through AI techniques (Mohammed et al., 2016; Rughani and Bhatt, 2017), automation (In de Braekt et al., 2016), and big data processing (Guarino, 2013).

* Corresponding author.
E-mail addresses: quan.le@ucd.ie (Q. Le), oisin.boydell@ucd.ie (O. Boydell), brian.macnamee@ucd.ie (B.M. Namee), mark.scanlon@ucd.ie (M. Scanlon).

## 1.1. Contribution of this work

The contribution of this work can be summarized as:

- An overview of existing techniques for malware analysis from a manual and automated perspective.
- An approach to enable malware classification by malware analysis non-experts, i.e., no expertise required on behalf of the user in reverse engineering/binary disassembly, assembly language, behavioral analysis, etc.
- Without using complex feature engineering, our deep learning model achieves a high accuracy of 98.2% in classifying raw binary files into one of 9 classes of malware. Our model takes 0.02 s to process one binary file in our experiments on a regular desktop workstation; this short processing time is of potential practical importance when applying the model in reality.
- Our one dimensional representation of a raw binary file is similar to the image representation of a raw binary file (Nataraj et al., 2011); but it is simpler, and it preserves the sequential order of the byte code in the binaries. The sequential representation makes it natural for us to apply the Convolutional Neural Network - Bi Long Short Term Memory architecture (CNN-BiLSTM) on top of it; helping us achieve better performance than using the CNN model alone.

## 2. Literature review/state of the art

There is a growing need for non-expert tools to perform digital evidence discovery and analysis (Sun, 2010; van de Weil et al., 2018). Due to the increasing delays in processing digital forensic evidence in law enforcement agencies throughout the world, there has been a focus in the digital forensic research and vendor communities in empowering the non-expert case detective to perform some preliminary analysis on the gathered evidence in a forensically sound manner (Lee et al., 2010). To this end, the Netherlands Forensic Institute (NFI) have implemented a Digital Forensics as a Service solution to expedite digital forensic processing (Casey et al., 2017). This system facilitates the case officer in uploading evidence to a private cloud-based system. Preliminary preprocessing takes place and the officer is able to browse the evidence to unearth potentially case-progressing information.

### 2.1. Digital forensic backlog

Storage capabilities are increasing exponentially while cybercrime related court cases are being dismissed. According to Ratnayake et al. (2014), the likelihood of a prosecution can be lessened due to the uncertainty in determining the age of a victim portrayed in a digital image. Their work considered a parallel challenge to age estimation which was to scan the sheer surface of disk drives. They are aware of the backlog that is eminent due to the lack of both relevant experts to analyze an offense and a laborious digital forensic process. Per Scanlon (2016), these factors will continuously influence the throughput of digital forensic laboratories; therefore, hinder digital forensic investigators in the future.

### 2.2. Machine learning for malware analysis

Machine learning offers the ability to reduce much of the manual effort required with traditional approaches to malware analysis, as well as increased accuracy in malware detection and classification. In the context of malware analysis, a machine learning model is trained on a dataset of existing labeled malware examples, with the labeling either in terms of *malicious* or *benign* in the case of binary classification, or in terms of the type or family of malware for multi-class classification. In either case, the model learns the differentiating features between the classes and so is able to infer, for a new and previously unseen example, whether it is malicious or benign, or which malware family it belongs to with a certain degree of accuracy.

Of course there are many different types and variations of machine learning algorithms and the training examples can be represented in many different ways, which all influence the classification accuracy of the resulting model. Research in the field generally involves the evaluation of different machine learning algorithms and approaches, in conjunction with different and novel types of features derived from the data. Many different approaches have been proposed and a comprehensive review of the literature is provided by both Ucci et al. (2017) and Gandotra et al. (2014).

In the next section, we focus specifically on approaches based on *deep learning* (a type of machine learning) as these are most related to our work. However, the types of features used and how they are extracted in the general context of machine learning for malware classification is also of key relevance. Machine learning reduces much of the manual effort required with traditional approaches to malware analysis by automatically learning to differentiate between malicious or benign, or different families of malware. However, the analysis and extraction of the features from the data, over which the machine learning model operates, still requires a high level of domain expertise in conjunction with complex and time consuming processes.

There are two families of features used in malware analysis: those which can be extracted from the static malware bytecode, and those which require the malware code to be executed (typically in a sandbox environment). Static features include information such as processor instructions, null terminated strings and other static resources contained in the code, static system library imports, and system API calls, etc. Features derived from executed code capture how the malware interacts within the wider operating system and network and can include dynamic system API calls and interactions with other system resources such as memory, storage and the wider network, e.g., connecting to external resources over the Internet.

Although dynamic features extracted from executed code are generally more time and computational resource consuming to extract than features from the static code, both cases require specialist tools and software environments – not to mention a high level of domain expertise required to understand and extract them. The core benefit of our approach, which we present in detail in the section 3, is that our deep learning model requires only the raw, static bytecode as input with no additional feature extraction or feature engineering.

Before moving on to review general deep learning approaches for malware classification in the next section, we first discuss two machine learning approaches which attempt to make use of the raw, static bytecode in a way which has some similarities to our work. Nataraj et al (2011) interpret the raw bytecode as greyscale image data where each byte represents a greyscale pixel, and they artificially wrap the byte sequence into a two dimensional array. They then treat the malware classification task as image classification by applying various feature extraction and feature engineering techniques from the image processing field, and use machine learning over these. Inspired by this approach, Ahmadi et al. (2016) use a similar representation of the data, and they evaluate this technique using the same dataset with which we evaluate our work, however they do not make use of deep learning. We provide a comparison of classification accuracy to our approach in the section 4.1. The application of image classification techniques to the malware domain however still requires the use of complex feature extraction procedures and domain expertise.

## 2.3. Deep learning for malware classification

Deep Learning ((LeCun et al., 2015), (Schmidhuber, 2015)) is a machine learning approach that has experienced a lot of interest over the last 5 years. Although artificial neural networks have been studied for decades, recent advances in computing power and increased data volumes have enabled the application of multi-layer neural networks (*deep* neural networks) to large training datasets, which has resulted in significant performance improvements over traditional machine learning techniques. Deep learning is now responsible for the state-of-the-art in many different machine learning tasks on different types of data, e.g., image classification (Hu et al., 2017) and natural language understanding and translation (Young et al., 2017). Malware classification has also attracted the attention of deep learning researchers.

The majority of deep learning approaches applied to malware classification involve training deep neural networks over the same types of extracted features on which traditional machine learning approaches are applied. These features require specialist knowledge and tools to generate and usually involve either the parsing or disassembly of the malware binary or running the malware in a sandbox environment and logging and analyzing the process execution and process memory, i.e., what the executed binary actually does (Schaefer et al., 2017). We survey various applications of deep learning to malware classification from the perspective of which types of data and features are used.

### 2.3.1. Features from static code

Saxe and Berlin (2015) present a deep feed forward neural network for binary malware classification that is trained on various features extracted from the static malware binary: system library imports, ASCII printable strings, metadata fields in the executable as well as sequences of bytes from the raw code. All these features require further processing and are then fed into a four layer feed forward network.

Hardy et al. (2016) propose the DL4MD framework (Deep Learning Framework for Intelligent Malware Detection), which is trained over API calls extracted from malware binaries. An API database is required to convert the call references from the format they are extracted from the code in to a 32-bit global ID representing the API call. These features are then used as input to a deep learning architecture based on stacked autoencoders.

Davis and Wolff (2015) discuss an approach whereby they apply a convolutional neural network for binary classification to disassembled malware byte code. The raw disassembled code is further processed to generate a more regularized set of features. For example, they extract the individual x86 processor instructions, which are variable length, and then apply padding or truncation to create fixed length features. They also parse the disassembled code to extract code imports, which they use to generate a further fixed length feature vector for each example.

All the aforementioned approaches require differing degrees of in-depth analysis of the disassembled code to extract domain specific features, which are then fed into various deep learning architectures. A key differentiator of our approach is that we do not require any domain specific analysis or parsing of the raw malware executable byte code. Our deep learning architecture requires no additional information as to the *meaning* of the raw data, or how it should be interpreted by the neural network. Although we do still need to normalize the length of the input data, as this is a basic requirement of the deep learning architecture we use, we do so at the entire raw malware file level and we use a context independent method to achieve this as described in the section 3.2.

Our methodology eliminates the need for complex feature engineering requiring expert domain knowledge and tools such as disassemblers, is not limited only to malware compiled for a specific processor or operating system, and the deep neural network is able to learn complex features directly from the data rather than being constrained to those engineered by human experts.

### 2.3.2. Features extracted from executed code

As well as deep learning based malware classification based on features from parsed and disassembled static malware code as summarized above, many approaches also make use of features derived from running the malware in a sandbox environment and analyzing the behavior of the running process. Although the key advantage of our methodology is that it only requires the raw malware byte code as input, we also include the following summary of these alternative approaches.

As with more traditional machine learning based malware classification, the use of system API calls logged from running malware processes are a popular source of input features. Dahl et al. trained neural networks of between one and three hidden layers on features generated from system API calls as well as null terminated strings extracted from the process memory (Dahl et al., 2013). A random projection technique was used to reduce the dimensionality of the features to that which was manageable by the neural network. Huang and Stokes (2016) propose an alternative deep learning architecture which uses similar features, however their model addresses multi-task learning in which the same deep learning architecture provides both a binary malware/benign classification as well as a classification of the malware type.

David and Netanyahu apply a *deep belief network* (DBN) to log files generated directly by a sandbox environment. This captures API calls as well as other events from the running malware as a sequential representation (David and Netanyahu, 2015). Similarly, Pascanu et al. (2015) apply a Recurrent Neural Network (RNN) to event streams of API calls, in this case encoded as 114 higher level events by a malware analysis engine. The use of an RNN captures the relationships of these events across time, and is similar in function to the LSTM component of our deep learning architecture. However we use it to capture the positional relationship of patterns within the static malware bytecode file rather than temporal relationships.

Kolosnjaji et al. (2016) propose a similar deep learning architecture to our methodology, which is also based on CNN and LSTM layers. However, the input data are sequences of system API calls extracted using the same sandbox environment as used by David and Netanyahu's approach discussed above 30. The CNN layers capture local sequences of API calls, whilst the LSTM layers model the relationships between these local sequences across time. In our approach, since we do not require the actual execution of the malware code, the CNN layers instead capture local sequences and patterns within the bytecode on a spatial level, and the LSTM layers model their longer distance relationships throughout the file.

Rather than using simple API call sequences, Tobiyama et al. (2016) use a more detailed representation of malware process behavior. They record details of each operation such as process name, ID, event name, path of current directory in which the operation is executed, etc. They then apply a RNN to construct a behavioral language model from this data, whose output is converted into *feature images*. A CNN is then trained over these feature images to produce binary malware/benign classifications. As with the previously outlined approaches that use features extracted from executing the malware code, the process required to collect the data is complex and time consuming. In this particular case, each malware or non-malware example was executed and analyzed for 100 min (5 min of logging, followed by 5 min interval and this was repeated 10 times). A complex sandbox environment setup was also needed, which is likely to have been another factor which resulted in a limited evaluation dataset being generated - only 81 malware and 69 benign examples.

In real world scenarios, malware defense systems that utilize machine learning based malware classification must be able to adapt to new variants and respond to new types of malware. If the approach requires complex, time and resource consuming processes to extract features required for the machine learning model, this will adversely impact the usefulness of the solution. This is a key motivation for our approach and so we focus on using only the static, raw malware bytecode with minimal data preprocessing.

Before we describe our methodology in detail in the next section, we will conclude our literature review with two approaches that are most similar to our methodology. Raff et al. (2017) describe a very similar motivation for their deep learning approach for malware classification - the need to remove the requirement for complex, manual feature engineering. Similar to our work, they focus on the raw malware bytecode and the application of deep learning techniques directly to this data. However, when faced with the challenge of how to work with such long sequences of bytes, they took a different approach which involved designing an atypical deep learning architecture that could handle such long input sequences. Our solution, on the other hand, is to simply use a generic data scaling approach (down sampling) as a pre-processing step, after which a more standard deep learning architecture can be applied. Although this approach, which by its nature reduces the detail in the data, might intuitively be thought of as resulting in drastically reduced classification accuracy, we show through evaluation that sufficient signal remains in the data for the deep learning network to exploit and achieve very high accuracy levels.

Finally, motivated by Ahmadi's work (Ahmadi et al., 2016), and with similarities to (Nataraj et al., 2011), Gibert (Gibert Llauradó, 2016) applied a CNN to malware bytecode represented as two dimensional greyscale images. A similar down sampling approach as we employed was applied to normalize the size of each sample to 32 x 32 pixels. The key differences with our approach is that we use the raw malware bytecode in its original one dimensional representation (we don't artificially wrap the byte sequence to create 2D representation), and we preserve more detail by down sampling the data to 10,000 bytes rather than 1024 (32 x 32). In terms of deep learning architectures, we utilize LSTM layers on top of CNN layers in order to capture relationships among local patterns across the entire malware sample. We used the same evaluation dataset and experimental setup as the work by Gilbert so we could directly compare approaches, and we observed a significant increase in classification accuracy with our approach which we present in more detail in Section 4.1.

## 3. Methodology

In this section, we describe our deep learning based approach for malware classification in detail, including the dataset we used for our experiments, data preprocessing, deep learning architectures, and experimental design.

### 3.1. Dataset

For our experiments, we used the malware data from the Microsoft Malware Classification Challenge (BIG, 2015) on Kaggle (Ronen et al., 2018).[1] Although the Kaggle challenge itself finished in 2015, the labeled training dataset of 10,868 samples is still available and represents a large collection of examples classified into malware classes, as shown in Table 1. As well as being able to use this data to both train and evaluate our own deep learning approaches, the Kaggle challenge still allows the submission of predictions for a separate unlabeled test set of 10,873 samples for evaluation.

**Table 1**
Number of examples for each malware class in the dataset.

| Malware Class | Number of Examples |
| --- | --- |
| Ramnit | 1541 |
| Lollipop | 2478 |
| Kelihos_ver3 | 2942 |
| Vundo | 475 |
| Simda | 42 |
| Tracur | 751 |
| Kelihos_ver1 | 398 |
| Obfuscator.ACY | 1228 |
| Gatak | 1013 |

Each labeled malware example consists of a raw hexadecimal representation of the file's binary content, without the PE header (to ensure sterility). In addition, a metadata representation is also provided, which includes details of function calls, embedded strings, etc., which were extracted using a disassembler tool. As the focus of our work is the application of deep learning techniques to classify malware based on the raw binary file content, we only consider the raw hexadecimal file representations, and convert it to its binary representation.

### 3.2. Data pre-processing

One of the benefits of deep learning over other machine learning techniques is it's ability to be applied over raw data without the need for manual and domain specific feature engineering. This is a key motivation for our work - the ability to efficiently classify malware without requiring specialist expertise and time consuming processes to identify and extract malware signatures. To parallelize the computation in training and testing the models efficiently, our deep learning approach requires that each file be a standard size, and in the case of malware the file size is highly variable, as shown in Fig. 1.

In addition to having the same size, from a computational perspective, our deep learning methods require that this size is constrained so as to keep the model training process practical using standard hardware. There are a number of options we could have taken to standardize the file size including padding and truncation, however we design our deep learning models to identify and detect common patterns and structure within the malware file data; hence we want to preserve the original structure as much as possible. To this end, we used a generic image scaling algorithm, where the file byte code is interpreted as a one dimensional 'image' and is scaled to a fixed target size. This is a type of lossy data compression. However, by using an image scaling algorithm, we aim to limit the distortion of spatial patterns present in the data. Compared to approaches of converting a malware binary file to a 2D image before doing classification, our approach is simpler since we do not have to make the decision about the height and width of the image. Also converting a binary file to a byte stream preserves the order of the binary code in the original file, and this sequential representation of the raw binary files makes it natural for us to apply a recurrent neural network architecture to it. In our experiments that follow, we scale each raw malware file to a size of 10,000 bytes using the OpenCV computer vision library (Bradski, 2000) - i.e. after the scaling one malware sample corresponds to one sequence of 10,000 1-byte values.

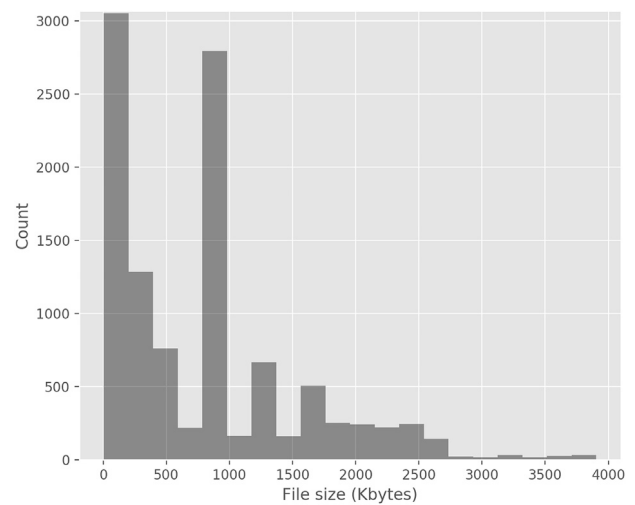

**Fig. 1.** Distribution of malware file size (in kilobytes) for the raw binary files in the training dataset.

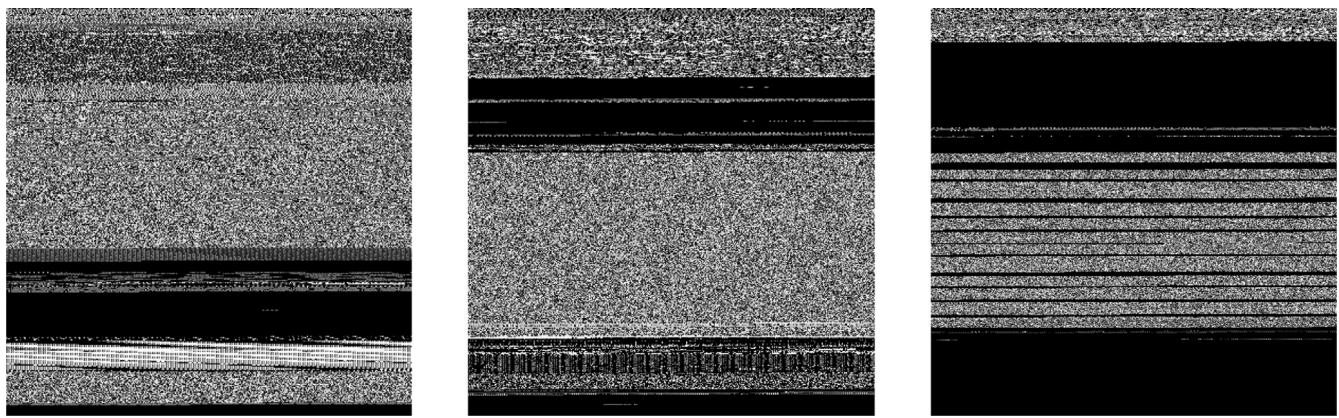

**Fig. 2.** Example malware binaries represented as greyscale images to aid visualization. Each byte is interpreted as a greyscale pixel, and the contiguous byte sequence is wrapped to produce a two dimensional image.

Fig. 2 shows a number of example malware files which have been scaled using this approach, and then represented as two dimensional greyscale images (one byte per pixel), where the images are wrapped into two dimensions purely for visualization purposes. The spatial patterns in the data both on a local scale and on a file level are visible and it is these raw features and patterns that our deep learning architecture is designed to exploit.

### 3.3. Deep learning architectures

We utilize different deep learning architectures for our experiments. We first apply multiple convolutional neural layers (CNNs) (LeCun et al., 1995) on the one dimensional sequential representation of the file. Since convolutional neural layers are shift invariant, this helps the models capture one dimensional spatial patterns of a malware class wherever they appear in the file.

On top of the convolutional layers, we apply two different approaches. In our first model, we connect the outputs of the convolutional layers to a dense layer, then to the output layer with a softmax activation to classify each input into one of the nine classes of malware, as shown in Fig. 3. This CNN-based approach classifies the one dimensional representation of the binary file using local patterns of each malware class, and is the dominant and very successful neural network architecture in image classification (Krizhevsky et al., 2012).

For the second and third models, we apply recurrent neural network layers, the Long Short Term Memory module (LSTM) (Hochreiter and Schmidhuber, 1997), on top of the convolutional layers, before feeding the output of the recurrent layer to the output layer to classify the input into one of the nine malware classes. Our rationale behind this approach is that since there are dependencies between different pieces of code in a binary file, a recurrent layer on top of the CNN layers will help to summarize the content of the whole file into one feature vector before feeding it to the output layer. In model two, CNN - UniLSTM, we apply one forward LSTM layer on top of the convolutional layer, where the connecting direction of the cells in the LSTM is from the beginning to the end of the file, as shown in Fig. 4. But since the dependency between code in a binary file does not go only in one direction, we design our third model, CNN-BiLSTM, where we connect the outputs of the convolutional layers to one forward LSTM layer and one backward

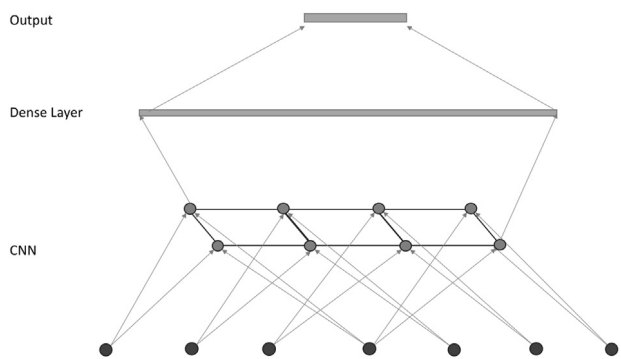

**Fig. 3.** Convolutional Neural Network (CNN) architecture.

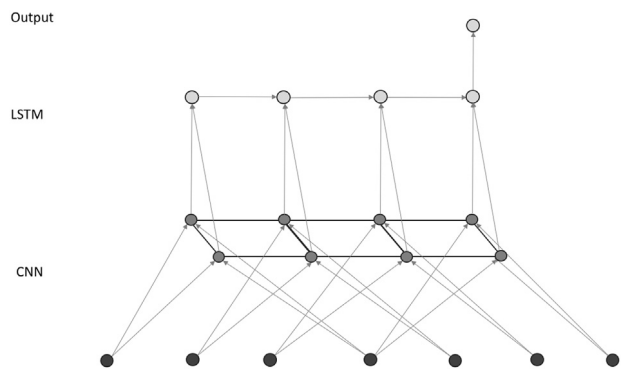

**Fig. 4.** Convolutional Neural Network plus Long Short Term Memory (CNN + LSTM) architecture.

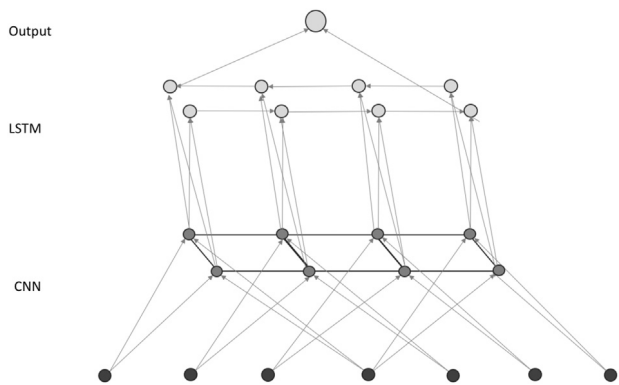

**Fig. 5.** Convolutional Neural Network plus bi-directional Long Short Term Memory (CNN + biLSTM) architecture.

LSTM layer. The outputs of the two LSTM layers are then concatenated and fed to the output layer, as can be seen in Fig. 5.

### 3.4. Experiment protocol

Since we only have the labels of the malware files on the training set of the Kaggle challenge, except for the final step of submitting the predictions on the test set to the Kaggle website each of the experiment results we report here are measured on this set of samples. For simplicity, we will refer to the training set of the Kaggle challenge as the main dataset.

After the preprocessing step we have 10,860 labeled samples in our dataset. Since this is a not a significantly large number, to achieve a more robust accuracy measure we use five fold cross-validation. The dataset is shuffled and divided into five equal parts, each with roughly the same class distribution as the main dataset. For a chosen deep learning configuration, we set each of the five parts as the left out part, train one model on the other 4 parts and record the predictions for samples in it. We then assemble the predictions for all five parts and use them to compute the performance of the chosen deep learning configuration.

The distribution of the classes in the dataset is highly imbalanced, with the number of samples per class ranging from 42 samples for the class Simda to 2942 samples for the class Kelihos_v3. Besides using the micro average classification accuracy to report the performance of a model, we also assess the performance of a model by its macro-averaged F1-score for each of the classes. The F1-score reports the performance of the model on any one class as the harmonic mean of the precision and recall on that class, and the macro average F1-score will treat the performance on each class of equal importance.

We take one additional step to address the class imbalance problem. In one training step of a deep learning model, a batch of chosen size, e.g. 64 samples, will be drawn from the training data, then the forward computation and the backward propagation is used to modify the weights of the model toward better performance.

The default sampling mode where all samples are drawn randomly from the training data will take samples mostly from the populous class, while likely missing samples from a rare class, such as Simda. To address this, in conjunction with using the default sampling procedure to generate data batches, we test a class rebalancing sampling approach, where for each batch we draw approximately the same number of samples from each class randomly. One batch of samples, of size $batch\_size \times sequence\_length$ is fed to the deep learning model without using the data normalization step.

In total, we have six deep learning configurations: each configuration is a combination of one of three deep learning architectures (CNN, CNN-UniLSTM, CNN-BiLSTM), and one of the two batch sample generating procedures in training the model (the default sampling mode, and the class rebalance sampling mode). All models have three convolutional layers, while the hyperparameters of a deep learning configuration, i.e., the number of nodes in each layer, is chosen through its performance in the cross-validation procedure.

To avoid overfitting, we use L2 regularization to constrain the weights of the convolutional layers, and dropout in the dense and LSTM layers. We choose the batch size to be 64. Other hyperparameters, e.g., the number of nodes in each layer, are chosen through the 5-fold cross-validation procedure.

Once the best deep learning configuration is chosen, we retrain the model on the whole training set, predict the labels for the malware files in the unlabeled test set, and submit them to the Kaggle website to get back the test set average log-loss - a low average log-loss correlates to a high classification accuracy.

## 4. Results and discussion

### 4.1. Results

Our final deep learning models' hyper-parameters are as follows. All models have three layers of convolutional layers with the rectified linear unit (ReLU) activation function; the number of filters at the three layers are 30, 50, and 90. For the CNN models, the outputs of the convolutional layers are connected to a dense layer of 256 units, then fed to the output layer. For the CNN with UniLSTM or CNN with BiLSTM models, we connect the outputs of the convolutional layers to one (UniLSTM) or two LSTM layers (BiLSTM), each LSTM layer has 128 hidden units; the outputs of the LSTM layers are then connected to the output layer. As described earlier, to complete a deep learning configuration each deep learning architecture (CNN, CNN-UniLSTM, CNN-BiLSTM) will be paired with one of the two data batch generators: the default sampling batch generator (DSBG), and the class rebalance batch generator (CRBG). The models are implemented using the Keras library with the Tensorflow backend.[2]

In the 5-fold cross-validation procedure, we train each model for 100 epochs on our Nvidia 1080 Ti GPU; the weights of the model are modified by the Adam optimization method (Kingma and Ba, 2014) to minimize the average log-loss criteria (i.e. the average cross-entropy criteria).

Table 2 reports the number of parameters and the training time for the six deep learning configurations. We report the average accuracy and the F1-score of different deep learning configurations in Table 3.

From the results, the CNN-BiLSTM with the class rebalance sampling batch generator configuration has the best F1-score and the best accuracy on validation data. As a result, we train our final model with this configuration on the entire training dataset, where 90% of the dataset is used to tune the weights of the model and the remaining 10% of the dataset is used as the validation data to choose the best model among the 100 epochs.

**Table 2**
The number of parameters and the training time for the specified deep learning Configurations.

| Configurations | No Params | Train time (m) |
|---|---|---|
| CNN - Def Sampl | 1,842,069 | 5.6 |
| CNN - Reb Sampl | 1,842,069 | 10.1 |
| CNN UniLSTM - Def Sampl | 155,669 | 32.1 |
| CNN UniLSTM - Reb Sampl | 155,669 | 55.1 |
| CNN BiLSTM - Def Sampl | 268,949 | 62.1 |
| CNN BiLSTM - Reb Sampl | 268,949 | 106.2 |

**Table 3**
Average accuracy and F1-score of different deep learning configurations using the 5-fold cross-validation procedure.

| Deep Learning Conf | Acc (%) | F1 (%) |
|---|---|---|
| CNN - Def Sampl | 95.1 | 92.14 |
| CNN - Reb Sampl | 95.8 | 92.14 |
| CNN UniLSTM - Def Sampl | 97.64 | 94.15 |
| CNN UniLSTM - Reb Sampl | 98.12 | 95.92 |
| CNN BiLSTM - Def Sampl | 97.91 | 95.52 |
| CNN BiLSTM - Reb Sampl | 98.20 | 96.05 |

**Table 4**
Time to pre-process and predict binary files in the test set with the final model.

| Stages | Total (s) | Average (s) |
|---|---|---|
| Convert bin file to 1D rep | 191.22 | 0.0176 |
| Prediction | 23.1 | 0.0021 |

Fig. 6 visualizes the loss and the accuracy on the training and validation data for the final model.

The final CNN-BiLSTM model achieves an average log-loss of 0.0762 on the validation data and a validation accuracy of 98.80%. Upon submitting the predictions of this model for the test malware files to Kaggle, we receive two average log-loss scores: a public score of 0.0655 calculated from 30% of the test dataset and a private score of 0.0774 calculated from 70% of the test dataset. These results align with the log-loss we obtained on the validation data, which means our final model generalizes well on new data.

Table 4 reports the times our final model takes to pre-process and predict the classes for the 10,873 test files. To simulate a real-life deployment situation, we load our final model onto a CPU (Intel Core i7 6850K) to do the predictions.

### 4.2. Discussion

Our experiments show that the one dimensional representation of the raw binary file is a good representation for the malware classification problem. It is very similar to the image representation of the malware raw binary file; however it is simpler, it preserves the sequential order of code in the raw binary file, and one does not have to make the decision about the ratio between the width and the height in the image representation.

Our use of the class rebalance sampling procedure helps to improve both the accuracy and the F1 score of all the CNN LSTM models (both the UniLSTM and BiLSTM models). We believe this improvement is due to the fact that the inclusion of samples of all classes in each batch gives the back propagation a better signal to tune the parameters of the models.

The best performance was achieved when training the CNN-BiLSTM with the class rebalance sampling procedure. Due to the sequential dependency when computing the cells in the LSTM layer, the CNN BiLSTM can not utilize the GPU as efficiently as the CNN model. With the same batch sampling procedure, training a CNN model is 10 times faster than training a CNN - BiLSTM model. On the other hand, the CNN-BiLSTM model uses 268,000 parameters while the CNN model uses 1.84 million parameters. When we use both models to predict the classes of raw binary files on the CPU, the CNN–BiLSTM model is only 1.5 times slower than the CNN model. The CNN - UniLSTM model trained with the class rebalance sampling procedure is a nice compromise; training it takes less time than training the CNN-BiLSTM model but it still achieves good performance.

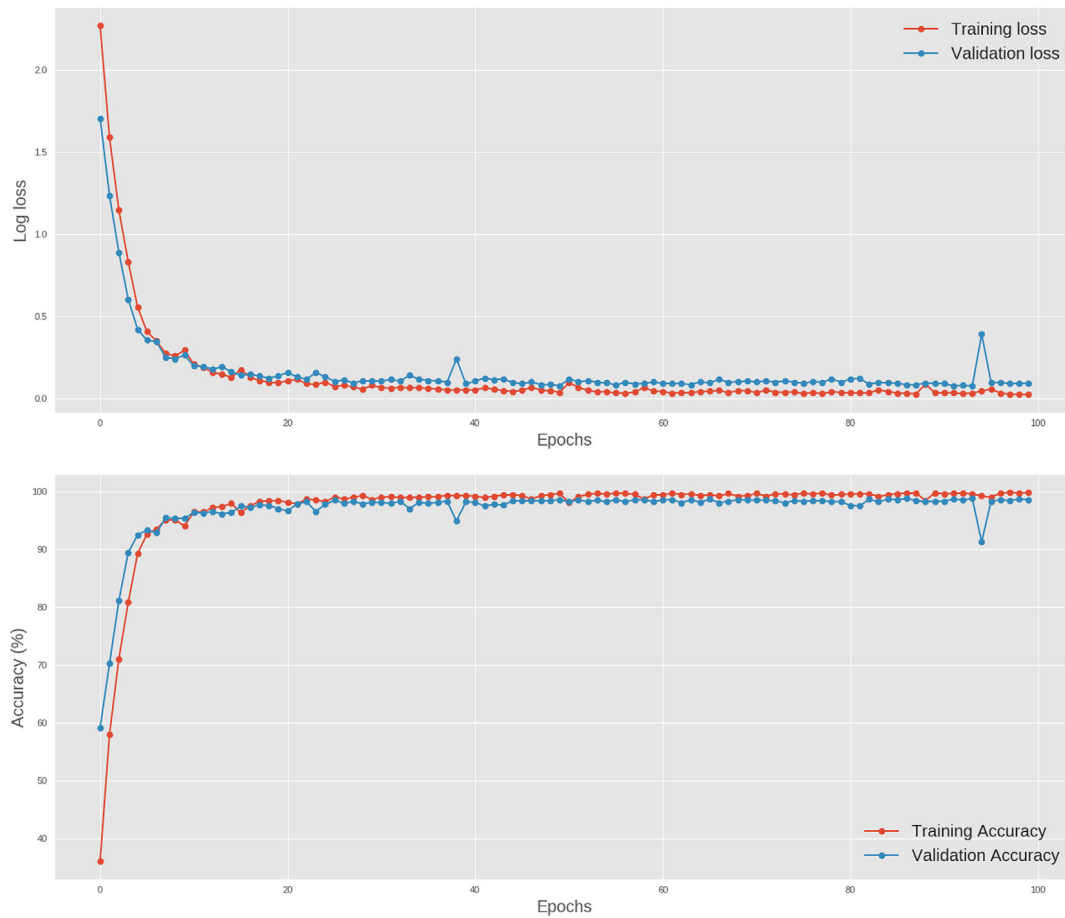

**Fig. 6.** Loss and accuracy on training the final model.

The results also show that adding another dependency direction in the binary code when going from using only the forward LSTM layer (CNN-UniLSTM model) to using both the forward and backward layer (CNN-BiLSTM model) helps improve the performance of the deep learning model. However the bigger jump in performance is achieved when we go from the CNN architecture to the CNN - LSTM architecture.

Ahmadi et al., 2016 also evaluate a machine learning-based approach to malware classification using the Kaggle dataset. Their feature engineering approach used a combination of different features extracted from the raw binary files and the disassembled files. One of them is the features extracted on the image representation of the raw binary files. Using the XGBoost classifier on these extracted features of the image representation they obtain the performance of 95.5% accuracy on the 5-fold cross-validation procedure, as shown in Table 4 of (Ahmadi et al., 2016). While our one dimensional representation of the raw binary file is similar to the image representation of raw binary file, our deep learning does not use feature extraction on top of it, and our best deep learning model obtains the accuracy of 98.2%, which is better than the previous feature engineering approach.

Another advantage of the deep learning approach is the time it takes to classify a new binary file. While training the models requires a GPU, the final model only needs to use a CPU to predict the malware class of a new binary file. Using our regular workstation with a 6 core i7-6850K Intel processor, training and testing files our final model takes on average 0.02 second to classify a binary file. This includes the time taken to convert a binary file to its one dimensional representation and the prediction time. As a comparison, two image feature extraction techniques in (Ahmadi et al., 2016) take on average of between 0.75 and 1.5 s for each binary file, as can be seen in Figure 8 in Ahmadi et al., 2016.

Gibert Llauradó (Gibert Llauradó, 2016) (Chapter 5) uses an approach similar to ours when using convolutional neural networks on the image representation of raw binary file. The CNN model they describe has 34.5 millions parameters; it has a public score of 0.1176 and a private score of 0.1348. Our CNN - BiLSTM model achieved a better performance with a public score of 0.0655 and a private score of 0.0774 while using 268,000 parameters.

## 5. Concluding remarks

Our deep learning approach achieves a high performance of 98.2% accuracy in the cross-validation procedure, and the final model has 98.8% accuracy based on the validation data. The appeal of the outlined deep learning approach for malware classification is two fold. Firstly, it does not require feature engineering, which is a big obstacle for researchers who are not familiar with the field. Secondly, the model takes a short time to classify the malware class of a binary file (0.02 s in our experiments), hence it is practical to use it in reality.

The results also show that the class rebalance batch sampling procedure could be used to address the class imbalance problem in the dataset. In practice, new malware families belonging the malware families recognized by the model will be found over time. For the deep learning approach, one could start from an available model and update it with new training data to improve its accuracy, thus the cost of retraining the model is small.

Our one dimensional representation of the raw binary has its limitations: it does not consider the semantics of the binary code in the raw binary file. However, as our experiments show, there are spatial patterns of each malware class in the raw binary files, and deep learning models could use them to predict the class of a malware file effectively. Gibert Llauradó (Gibert Llauradó, 2016) shows that one could apply deep learning on the disassembled files successfully, it shows that there are merits in considering the semantic meaning of each byte − even if the reverse engineering step is not conducted through disassembling the raw binary files.

## Future work

For future work, we would like to test our deep learning approach on bigger datasets with more malware classes. One approach is to preserve the semantic meaning of each byte in the raw binary file in the preprocessing step, though this approach means we need a suitable way to compress a large binary file (approximately 60 Mbytes) to a small size without losing the semantic meaning of the bytes in the final representation. Another useful feature would be to modify our deep learning model so it could detect if the new binary file belongs to one of the available classes or belongs to a new malware class. Finally, we could apply more complex deep learning architectures to attain better performance, for example we could add residual modules He et al., 2016 42 to the model to alleviate the vanishing gradient problem.